**Pauli-Spin-Blockade Transport through a Silicon Double Quantum Dot**

H.W. Liu,[1,2,3] T. Fujisawa,[1] Y. Ono,[1] H. Inokawa,[1,a] A. Fujiwara,[1] K. Takashina,[1] and Y. Hirayama [1,2,4]

[1]NTT Basic Research Laboratories, NTT Corporation, 3-1 Morinosato-Wakamiya, Atsugi, Japan

[2]SORST-JST, Chiyoda, Tokyo 102-0075, Japan

[3]National Laboratory of Superhard Materials, Institute of Atomic and Molecular Physics, Jilin University, Changchun 130012, P.R. China

[4]Department of Physics, Tohoku University, Sendai, Miyagi 980-8578, Japan

[a] Present address: Research Institute of Electronics, Shizuoka University, 3-5-1 Jyouhoku, Hamamatsu, Japan

We present measurements of resonant tunneling through discrete energy levels of a silicon double quantum dot formed in a thin silicon-on-insulator layer. In the absence of piezoelectric phonon coupling, spontaneous phonon emission with deformation-potential coupling accounts for inelastic tunneling through the ground states of the two dots. Such transport measurements enable us to observe a Pauli spin blockade due to effective two-electron spin-triplet correlations, evident in a distinct bias-polarity dependence of resonant tunneling through the ground states. The blockade is lifted by the excited-state resonance by virtue of efficient phonon emission between the ground states. Our experiment demonstrates considerable potential for investigating silicon-based spin dynamics and spin-based quantum information processing.

Electron spins in quantum dots are attracting increasing attention for the study of spin dynamics and quantum information devices. Up to now, extensive investigations have been carried out on electron spins of GaAs quantum dots, in which the electronic states are well understood.[1-3] However, alternative materials such as silicon (Si) exhibit highly advantageous spin properties. In Si, extremely long spin relaxation and coherence times are expected owing to it having desirable material characteristics: weak spin-orbit coupling, predominantly spin-zero nuclei, and the absence of piezoelectric electron-phonon coupling.[4,5] This, together with advanced metal-oxide-semiconductor (MOS) technology, makes spins in Si dots more attractive for building qubits (quantum bits). Moreover, the multiple valleys in the conduction band, large Zeeman effect (Lande g factor of ∼ 2), and anisotropic effective mass have been predicted theoretically to offer distinct spin characteristics.[6,7]

So far, successful mapping of spins in Si has been achieved in a *single* dot where the magnetic field dependence of energy levels is dominated by the Zeeman shift.[8] Several studies addressed electron transport and/or charge sensing measurements on Si double quantum dots (DQDs),[9-11] in which they demonstrated controllable charge states. Further evidences of spin dependent transport through the DQDs are required for investigating spin lifetime, coherence, and entanglement in Si nanostructures. A mandatory demand for such investigations is that quantized energy levels in each dot should be sufficiently well resolved so that electron transport measurements can be made on resonant (elastic) tunneling through discrete levels of the two dots, which, in turn, would provide a powerful way to probe the nature of their electronic (spin) states. Since the effective mass in Si is relatively large, each of the two dots needs to be very small to achieve well-separated quantized levels.

In this work, we present a clear evidence of resonant tunneling transport in an intrinsic Si DQD. The effective two-electron DQD is formed in multi-gated MOS structures with a thin silicon-on-insulator (SOI) layer. Two small dots with sufficient quantized level separation are created in this thin SOI layer. Appropriate electrostatic interdot coupling of the DQD has enabled us to demonstrate elastic tunneling as well as inelastic tunneling (energy dissipation) arising from spontaneous phonon emission with deformation-potential coupling, which is in contrast to the predominantly piezoelectric coupling in GaAs DQDs. Most importantly, we have succeeded in clearly observing a Pauli spin blockade in which electron transport is suppressed for the two-electron spin-triplet states,[12,13] as is evident from an obvious bias-polarity dependence of resonant tunneling through the ground states (GSs) of the two dots. The blockade is lifted through aligned excited states (ESs) since the phonon emission between the off-resonant GSs is efficient. Our findings represent a significant step forward for coherently manipulating single electron spin and the exchange of two neighboring spins with the expected long coherence time as well as for elucidating spin dynamics in Si nanostructures.

Figure 1(a) shows a schematic cross section of the multi-gated MOS structures with a (100) SIMOX (*s*eparation by *im*plantation of *ox*ygen) SOI layer.[14] A top-view scanning electron microscopy (SEM) image of the device taken before the upper poly-Si gate (UG) had been fabricated is shown in Fig. 1(b). When a positive voltage $V_{UG}$ is applied to the UG, electrons accumulate at the interface between the gate oxide and the SOI layer. Since our previous measurements indicate that the valley degeneracy is lifted in such a two-dimensional electron system,[15,16] the valley degree of freedom is not discussed unless noted otherwise in this paper. The lower poly-Si gates ($LG_1$, $LG_2$ and $LG_3$) are used to form tunneling barriers. The SOI layer thickness plays an important role in forming the dots. When it is thick (20 nm), each lower gate defines a single barrier and single or double dots can be formed intentionally using two or three lower gates.[11] However, quantized energy levels in such dots with relatively large size are not shown. When a thin SOI layer (~ 10 nm) is used, a tunneling barrier under a lower gate is often accompanied by a small dot thought to arise from the additional confinement of the buried oxide and the associated thickness variations of the SOI nanowire. Two small dots involving well-separated energy levels can then be readily formed between two neighboring lower gates. All experiments were performed on such a small DQD [white circles in Fig. 1(b)] using a dc current measurement at zero magnetic field (unless mentioned otherwise) in a dilution refrigerator with a base temperature of 100 mK.

Current through the DQD (*I*) as a function of lower-gate voltages $V_{LG1}$ and $V_{LG2}$, shown in Fig. 1(c), was measured at source-drain voltage $V_D = 1$ mV and $V_{UG} = 5$V. The equilibrium electron numbers (n,m) in dots L and R can be controlled by $V_{LG1}$ and $V_{LG2}$. Transport through the DQD will be affected by electrostatic and tunneling coupling,[17,18] both of which may lead to a peak splitting. In our case, the observed hexagonal charge domains outlined by current peaks are consistent with a well-accepted capacitance model[17] with large electrostatic coupling energy U (~ 1.7 meV) and negligible tunneling coupling. Although each dot might contain around 10 electrons as described below, the labeled effective electron numbers in the range from 0 to 2 explain the following experiments well. A fine scan of the rectangular region at positive bias $V_D = 1.1$ mV is shown in Fig. 2(a), where two triangular conductive regions representing sequential tunneling transport are clearly observed (marked by solid gray lines).[17] Fig. 2(b) shows the excitation spectrum of dot L where cotunneling current $I$ is traced at different $V_D$ by sweeping $V_{LG1}$ along the arrow α in Fig. 2(a). The current increases stepwise when the electrochemical potential $E_i$ ($i$ = g for the GS; $i$ = e1, e2 and e1' for the ESs) of the (2,1) and (1,1) charge states in dot L coincides with the Fermi energy $\mu_{S(D)}$ of the leads.[19] An analysis of the excitation spectrum gave a charging energy $E_C \sim 16$ meV (corresponding to a dot radius $r \sim 12$ nm) and a level spacing $\Delta \sim 0.7$ meV. The electron numbers of $n \sim 10$ were estimated using the measured electron areal density $\sim 2.5 \times 10^{12}$ cm$^{-2}$ of the accumulated electron layer. The parameters of dot R ($E_C \sim 18$ meV, $r \sim 11$ nm, $\Delta \sim 0.8$ meV, $n \sim 9$) were obtained in a similar way [current is traced by sweeping $V_{LG2}$ along the arrow β in Fig. 2(a)].

In Fig. 2(a), current peaks A and B correspond to resonant tunneling through the GSs of the two dots and peaks C and D arise from resonant tunneling through the ES of dot R (discussed below). Besides the elastic current, inelastic tunneling current is also seen inside the triangular conductive regions. The current profile as a function of the energy difference ε between the two-electron GSs is plotted in Fig. 2(c). In addition to resonant peaks B and D, a shoulder band, accentuated by a hatched pattern, is superimposed on the right side of peak B at $V_D = 0.7$ mV before the onset of peak D at $V_D = 0.8$ mV. This asymmetric structure is attributed to inelastic transitions (phonon emission)[20,21] between the two GSs, as shown in the inset. The phonon emission rate $\Gamma_{ph} \approx 1.3 \times 10^9$ Hz was estimated from the inelastic current of $\sim 0.2$ nA at $\varepsilon \sim 0.3$ meV. In the absence of piezoelectric electron-phonon coupling (piezo), theories predict that the deformation potential coupling (def) accounts for the phonon emission rate in Si dots.[22,23] We compared the obtained $\Gamma_{ph}$ (◇) with the calculation one (solid line) in Fig. 2(d). For simplicity, the phonon emission from the antibonding state to the bonding state at the resonant condition of $\varepsilon = 0$ (corresponding energy diagram is plotted in the inset) was calculated by considering the isotropic deformation coupling of Si [23] and the geometric parameters of our Si dots. The $\Gamma_{ph}$ and the calculation are found to be of the same order of magnitude, although the inelastic rate at finite ε is expected to be smaller than that at $\varepsilon = 0$ because of the small wavefunction overlapping. This crude comparison suggests that the phonon emission in Si DQDs is relatively strong at this energy scale. The $\Gamma_{ph}$ in a GaAs DQD was also calculated by taking both deformation and piezoelectric coupling into account (dashed line), which is comparable to that measured during charge qubit operations in GaAs DQDs (solid circle, data from Ref. 24). As clearly shown in Fig. 2(d), the $\Gamma_{ph}$ in Si DQDs becomes several orders of magnitude less than that in GaAs DQDs in the energy range of 1~10 μeV where electrical control of qubit operation is expected, suggesting the suitability of Si DQDs for qubit demonstration. However, further studies such as spin-relaxation measurements are needed to prove it.

At negative bias, electron transport through the DQD is different from the positive bias case. Fig. 3(a) shows a fine scan of the rectangular region in Fig. 1(c) at $V_D$ = -1.1 mV, where sequential tunneling is expected in the triangular regions marked by solid gray lines. Peaks G and H showing relatively strong current are attributed to resonant tunneling through the ES of dot L as discussed below. Resonant tunneling through the GSs of the two dots, however, is strongly suppressed except at the outmost point of the upper triangle (point E). For clarity, the current evolutions as a function of ε are shown in Fig. 3(b) at various $V_D$. At small bias ($|V_D| \lesssim 0.1$ mV), peaks at ε = 0 (⊕) denote a linear transport through the two GSs. As the bias increases, current at ε = 0 is suppressed at negative bias and another peak H at ε ~ -0.5 meV is observed at $V_D$ < -0.5 mV. A Pauli spin blockade accounts for these transport features, assuming that one or two electrons in DQDs contribute to the transport and other spin-paired electrons can be neglected. When an electron (say, spin-down) occupies the left dot [(1,0) charge state], another spin-down electron is not allowed to tunnel from the right lead to the left lead. The electron is blocked in a (1,1) triplet state, $(1,1)_T$, and can not enter a singlet $(2,0)_S$ due to the Pauli principle [inset to Fig.3(c)]. A blockade of this type can not occur when electrons proceed from the left to the right at positive bias. As a result, a clear current rectification is shown in Fig. 3(c), where an $I$-$V_D$ plot is taken at the triple point.[17] When the (1,1) state nears the Fermi levels in the leads, a direct tunneling from $(1,1)_T$ to (2,1) favors another electron occupying a singlet $(1,1)_S$ and then lifts the blockade. Such spin exchange processes contribute to the linear conductance at $|V_D|$ < 0.1 mV in Fig. 3(c) and the relatively strong current peak at point E in Fig. 3(a). Another current peak is also expected at point F but is not present because of the nearly quenched left tunneling barrier. In addition, the leakage current ~ 4 pA in the blockade region of Fig. 3(c) ($V_D$ < -0.3 mV) is attributed to cotunneling from $(1,1)_T$ to $(1,1)_S$ via the (2,1) charge state.[12,25]

The exchange splitting of spatially separated (1,1) charge states is important for spin qubit operations,[2,26] however, it is too small to be measured in our device. Here we take advantage of the spin blockade to detect the spin splitting of (2,0) charge states. The large current at peak H in Fig. 3(a) suggests that the corresponding ES is a triplet $(2,0)_T$, at which resonant tunneling from $(1,1)_T$ to $(2,0)_T$ lifts the blockade [Fig. 3(d), lower right inset]. Note that, the blockade is lifted by virtue of efficient phonon emission from $(1,1)_S$ to $(2,0)_S$. The singlet-triplet splitting $\Delta_{ST}$ between the $(2,0)_S$ and $(2,0)_T$ states [energy spacing between $\oplus$ and H in Figs. 3(a) and 3(b)] corresponds to the orbital energy difference compensated by the exchange energy of the two electrons in dot L. Fig. 3(d) shows $\Delta_{ST}$ as a function of the magnetic field $B$ perpendicular to the substrate. The $\Delta_{ST}$ decreases with increasing $B$ and is expected to approach zero at ~ 4 T. An analogous characteristic is also observed in GaAs DQDs.[27] The $B$-dependent experiment further supports the presence of the Pauli spin blockade in our Si DQDs. In addition, peak D in Fig. 2(a) at positive bias represents an ES in dot R as mentioned above. The large current peak indicates that the ES is a singlet $(1,1)_{S'}$ and electrons in $(2,0)_S$ are thus allowed to enter [Fig. 3(d), upper right inset]. As shown in Fig. 3(d), the energy spacing $\Delta_{SS'}$ between the $(1,1)_S$ and $(1,1)_{S'}$ states (single particle excitation in dot R) is almost independent of $B$. These two singlet states are attributed to different orbits in dot R or to valleys of the conduction band.[15,16] Further studies would reveal the interplay between the orbit, valley, and spin degrees of freedom in Si dots. Note that, similar spin blockade phenomenon has also been demonstrated in Si/SiGe DQDs recently.[28]

In summary, a Pauli spin blockade has been successfully observed and applied to identify spin singlet and triplet states in an effective two-electron Si DQD formed in a thin SOI layer. The blockade is lifted by the excited-triplet resonance in virtue of efficient phonon emission through the off-resonant ground singlet states. The phonon emission dominated by deformation-potential coupling is relatively strong in the transport spectrum but is estimated to be weak in the qubit-operation energy range. These results, together with on-chip electron spin resonance techniques,[3] demonstrate the feasibility of spin qubits with the expected long coherence time. Moreover, we can take advantage of the Pauli spin blockade to probe the hyperfine coupling to 4.7% abundant $^{29}$Si nuclei or a few doping nuclei in Si dots as well as the inter(intra)-valley-dependent exchange interaction between two neighboring spins.[7] Accordingly, our experiment opens up quite appealing perspectives for Si-based spin dynamics study and spin-based quantum information processing.

We thank T. Ota for useful discussions. This work was partially supported by MEXT KAKENHI (16206038 and 16206003) and SCOPE. H.W. Liu also thanks the National Natural Science Foundation of China under Grant No. 10574055.

**FIG. 1**. (Color online) (a) Schematic cross-section of a Si MOS structure with three lower gates $LG_1$, $LG_2$, and $LG_3$. A bias voltage was applied between the source (S) and the drain (D). Thicknesses of the buried oxide, the gate oxide, and the oxide around the lower gates are 400, 35, and 30 nm, respectively. The width of the lower gate is 15 nm and the gate spacing is 85 nm. (b) Top-view SEM of the device before fabricating the upper gate UG. The Si wire width is 40 nm. Two dots L and R are formed between gates $LG_1$ and $LG_2$ (marked by white circles). (c) Current $I$ through the DQD as a function of $V_{LG1}$ and $V_{LG2}$. Hexagonal charge domains are outlined by dashed lines, where (n,m) denotes the effective electron numbers in dots L and R.

**FIG. 2**. (Color online) (a) Fine-scan of the rectangular region in Fig. 1(c) at $V_D$ = 1.1 mV. Solid and dashed gray lines outline the sequential tunneling and cotunneling regions, respectively. (b) Cotunneling current $I$ as a function of $V_{LG1}$ swept along the arrow α in (a). Each trace is shifted for clarity. (c) Current $I$ through the DQD as a function of the energy offset ε. The hatched pattern denotes inelastic tunneling through the DQD. The corresponding energy diagram and localized wavefunctions at ε > 0 are shown in the inset. (d) Dependence of phonon-emission rate $\Gamma_{ph}$ on the energy E. The calculation was done for the spontaneous phonon emission from the anti-bonding state to the bonding state at ε = 0 (corresponding energy diagram and wavefunctions shown in the inset) for GaAs (dashed line) and Si (solid line) DQDs. We consider Gaussian-shaped dots (radius R = 25 nm for GaAs and 12 nm for Si) separated by 2R. Deformation coupling (Ξ = 7 eV) and piezoelectric coupling ($e_{14}$ = 0.16 C/m$^2$) for GaAs and isotopic deformation coupling (Ξ = 3.3 eV) for Si were used.[23] Data from this work for Si DQDs (◇) and from Ref. 24 for GaAs DQDs (●) are plotted.

**FIG. 3**. (Color) (a) Fine-scan of the rectangular region in Fig. 1(c) at $V_D$ = -1.1 mV. (b) Current $I$ through the DQD as a function of ε at various $V_D$. Each trace is shifted for clarity. (c) $I$-$V_D$ plot traced at the triple point in the linear transport regime. Energy diagram denoting the Pauli spin blockade is plotted in the inset. (d) Energy spacing $\Delta_{ST}$ and $\Delta_{SS'}$ as functions of magnetic field $B$ perpendicular to the substrate. Corresponding energy diagrams are shown in the lower right and upper right insets, respectively.

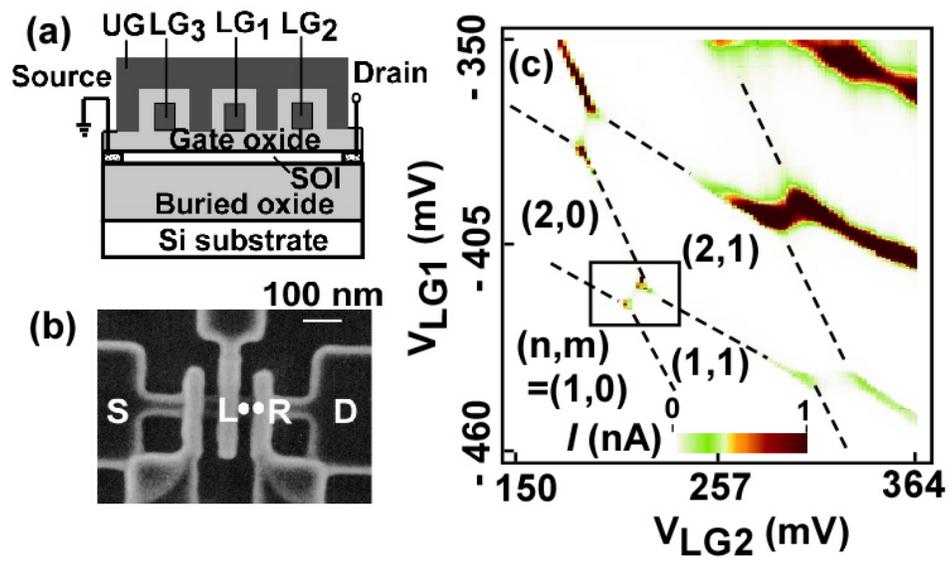

**FIG. 1.**
Liu HW

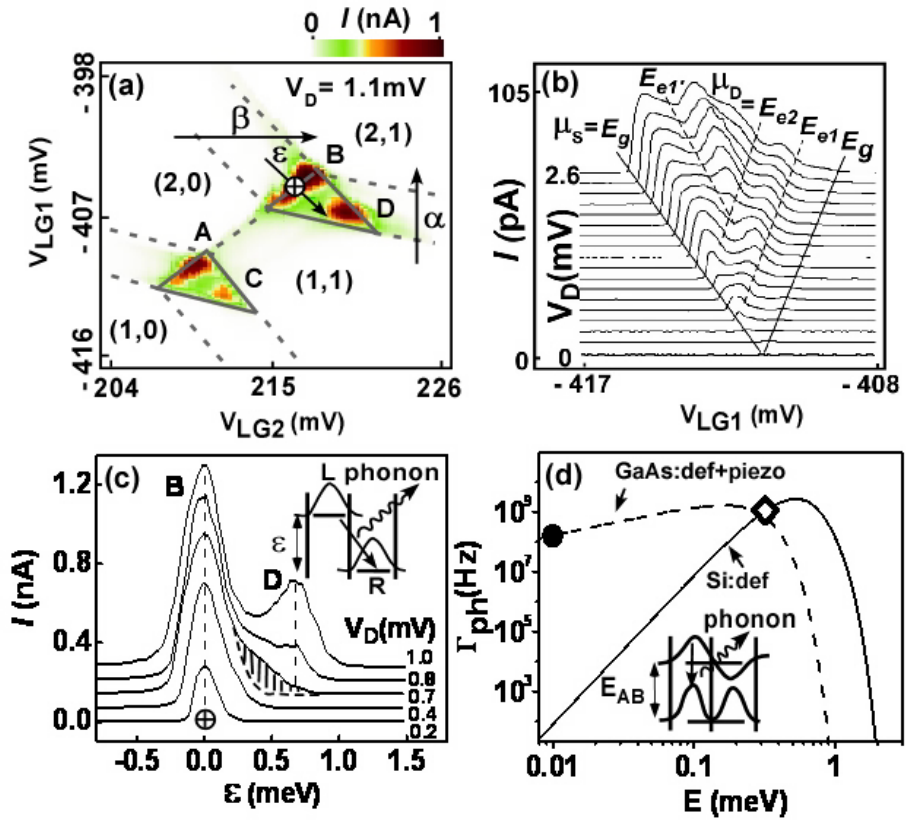

**FIG. 2.**
Liu HW

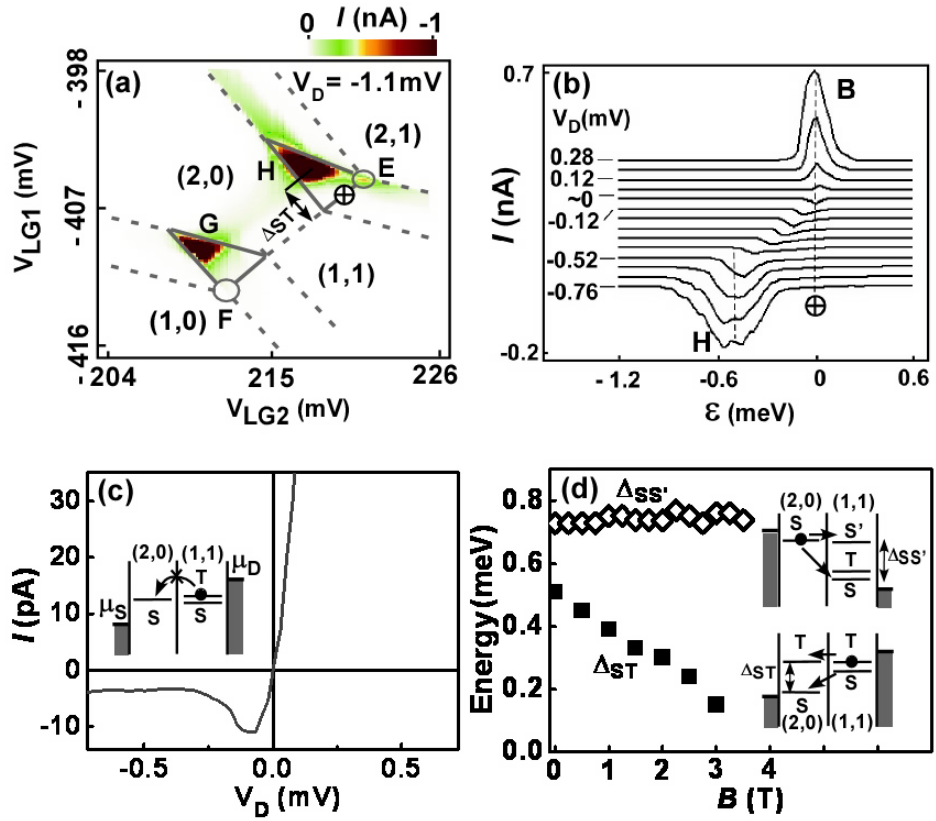

**FIG. 3.**
Liu HW